\documentclass[aps, prd, reprint, twocolumn, superscriptaddress]{revtex4-1}
\usepackage{epsfig}
\usepackage{graphicx}
\usepackage{color}
\usepackage{bm}
\usepackage{amssymb,amsmath,euscript,esint,mathtools}
\usepackage{times}
\usepackage{amsthm}
\usepackage{amsfonts}
\usepackage[english]{babel}
\usepackage{epstopdf}
\usepackage{multirow,makecell,array}
\usepackage{cancel}
\usepackage{float}
\usepackage[dvipsnames]{xcolor}
\usepackage[colorlinks=true,linkcolor=blue,citecolor=blue,urlcolor=blue]{hyperref}
\begin{document}
\newcommand{\Br}[1]{(\ref{#1})}
\newcommand{\Eq}[1]{Eq.~(\ref{#1})}
\newcommand{\frc}[2]{\raisebox{1pt}{$#1$}/\raisebox{-1pt}{$#2$}}
\newcommand{\frcc}[2]{\raisebox{0.3pt}{$#1$}/\raisebox{-0.3pt}{$#2$}}
\newcommand{\frccc}[2]{\raisebox{1pt}{$#1$}\big/\raisebox{-1pt}{$#2$}}
\newcommand{\RNumb}[1]{\uppercase\expandafter{\romannumeral #1\relax}}

\title{$\mathcal{T}$, $\mathcal{P}$-odd electron-nucleon interaction via a Higgs-boson exchange at the quark-gluon level}
\author{D. V. Chubukov}
\email{dmitrybeat@gmail.com}
\affiliation{School of Physics and Engineering, ITMO University, Kronverkskiy 49, St. Petersburg 197101, Russia}
\author{I. A. Aleksandrov}
\affiliation{Department of Physics, St. Petersburg State University, 7/9 Universitetskaya Naberezhnaya, St. Petersburg 199034, Russia}
\affiliation{Ioffe Institute, Politekhnicheskaya Street 26, St. Petersburg 194021, Russia}

\begin{abstract}
The electron pseudoscalar -- nucleon scalar $\mathcal{T}$,~$\mathcal{P}$-odd interaction within the framework of the Standard Model is considered as an exchange mediated by the Higgs boson at the quark-gluon level. $\mathcal{CP}$ violation is introduced through a phase factor in the Cabibbo-Kobayashi-Maskawa matrix, which characterizes the flavor structure of the quark loop.  We explicitly demonstrate that the total three-loop contribution vanishes for the Higgs boson exchange mechanism. A nonzero contribution appears at the four-loop quark-gluon level only with the insertion of an additional gluon line between the quarks. The estimate of the $\mathcal{CP}$-odd electron-nucleon interaction coupling constant after reduction of divergencies of the quark loop integral using the Glashow-Iliopoulos-Maiani (GIM) mechanism yields a value $ \sim10^{-28}$. The final result, expressed in terms of the equivalent electron electric dipole moment ($e$EDM), is $10^{-48} \,\, e\text{cm}$, and its order of magnitude coincides with that of the GIM estimate of the $e$EDM at the quark-gluon level. Our parametric estimate of the $e$EDM differs from previous studies by a factor that depends on the mass of the $b$ quark.
\end{abstract}
 
\maketitle

\section{Introduction}\label{sec:intro}

A nearly 75-year history of the search for $\mathcal{T}$-noninvariant ($\mathcal{T}$ is the time reversal) interactions in nature began with the paper~\cite{Pur50} which first discussed the existence of the neutron electric dipole moment (EDM) and proposed a magnetic resonance method to observe it. The existence of the EDM for any not truly neutral elementary particle, i.e., not identical to its antiparticle, such as leptons, quarks, and vector bosons, implies both space parity $\mathcal{P}$ and $\mathcal{T}$ violation. The same applies to closed systems of such particles, including nucleons, nuclei, atoms, and molecules. Due to the fundamental $\mathcal{C}\mathcal{P}\mathcal{T}$ conservation theorem, $\mathcal{T}$ violation is equivalent to the combined parity $\mathcal{C}\mathcal{P}$ violation ($\mathcal{C}$ is the charge conjugation). From the same theorem, it follows that $\mathcal{T}$,~$\mathcal{P}$ violation implies $\mathcal{C}$ conservation, so that all the $\mathcal{T}$, $\mathcal{P}$-odd effects should be the same for particles and antiparticles, atoms and antiatoms, etc., in general, for matter and antimatter. $\mathcal{C}\mathcal{P}$ violation was discovered in the decays of neutral $K$ mesons~\cite{Chris64}, and later of some other exotic mesons~\cite{BABAR01,Belle01, BABAR04, Belle04}. The first direct observation of $\mathcal{T}$ noninvariance was performed again in meson physics~\cite{Lees12}. However, the existence of EDMs may mean the universality of $\mathcal{T}$-odd interactions in nature and the possibility of time reversal for separate particles in the corresponding processes (see, e.g., Ref.~\cite{Zal22}).

A search for the electron EDM ($e$EDM) began with the paper~\cite{Sal58}, where it was suggested that the observation of an atomic paramagnetic EDM (i.e. connected to unpaired electron spin) would indicate the manifestation of the $e$EDM. Later, in Refs.~\cite{San65,Flam76}, it was discovered that the $e$EDM is significantly enhanced in heavy atoms compared to the free $e$EDM. In Ref.~\cite{San67}, it was theoretically demonstrated that a strong enhancement of nuclear EDMs can be observed in heavy diatomic molecules with closed electron shells. An enhancement of the $\mathcal{P}$-odd effects in heavy diatomic molecules with open shells that is even stronger than in atoms was predicted in Refs.~\cite{Lab77,Lab78}. This enhancement arises from the unique property of diatomic molecules known as the $\Lambda$-doubling effect, i.e., the splitting of every electron energy level (including the ground state) into two very close components with opposite space parities. The quantum number $\Lambda$ represents the projection of the total electron orbital angular momentum on the molecular axis. In heavy molecules, where parity-violating effects are most pronounced, $\Lambda$ should be replaced by $\Omega$, which is a sum of projections of the total orbital and total spin momenta of an electron system. This effect is absent in molecules with closed electron shells. Similar effects, as well as the $\mathcal{T}$,~$\mathcal{P}$-odd effects associated with the $e$EDM were considered in Refs.~\cite{San75,Sush78,Gor79}. In Refs.~\cite{San75,Gor79}, $\mathcal{T}$, $\mathcal{P}$-odd electron-nucleus interaction was introduced for the first time. In Ref.~\cite{Gor79}, it was also demonstrated that the $e$EDM effect and the $\mathcal{T}$,~$\mathcal{P}$-odd electron-nucleus interaction effect cannot be distinguished in any experiment with a particular atom or molecule. Later, in Ref.~\cite{Bon15}, it was shown that the two effects can be distinguished in a series of experiments with highly charged ions due to their different dependence on the nuclear charge $Z$.

The extraction of $\mathcal{T}$,~$\mathcal{P}$-odd parameters ($e$EDM values) from experimental data requires accurate theoretical calculations, which are especially involved for molecules. A general calculation scheme was developed in Refs.~\cite{Dmit87,Dmit92}. Later, this scheme was modified, further developed, and successfully applied to the analysis of the $\mathcal{T}$,~$\mathcal{P}$-odd effects (enhancement coefficients) in many diatomic molecules (see, for example, Refs.~\cite{Mos98,Is05,Tit06,Skrip09,Skrip16}). A semiempirical method for computing the $\mathcal{P}$- and $\mathcal{T}$,~$\mathcal{P}$-odd effects in diatomic molecules was developed in Refs.~\cite{Koz95,Koz97}. Other \textit{ab initio} calculations of $\mathcal{T}$,~$\mathcal{P}$-odd effects in diatomic molecules can be found, e.g., in Refs.~\cite{Quin98,Par98}.

The most stringent constraints on the $e$EDM were established in experiments with the Tl atom ($d_e<1.6\times 10^{-27}$~$e$cm~\cite{Reg02}), YbF molecule ($d_e<1.05\times 10^{-27}$~$e$cm~\cite{Hud11}), ThO molecule ($d_e< 0.87 \times 10^{-28}$~$e$cm~\cite{ACME13}, $d_e<1.1\times 10^{-29}$~$e$cm~\cite{ACME18}), and HfF$^+$ molecular ion ($d_e<1.3\times 10^{-28}$~$e$cm~\cite{Cair17}, $d_e<4.1\times 10^{-30}$~$e$cm~\cite{JILA23}). Here $e$cm stands for the elementary charge times centimeter. Calculations of the enhancement coefficients for the Tl atom were performed in Refs.~\cite{Liu92,Dzuba09,Por12,Chub18}, for the YbF molecule in Refs.~\cite{Quiney:98, Parpia:98, Mos98}, for the ThO molecule in Refs.~\cite{Skripnikov:13c,Skripnikov:15a,Skripnikov:16b,Fleig:16},  and for the HfF$^+$ molecular ion in Refs.~\cite{Petrov:07a,Skripnikov:17c, Fleig:17, Petrov:18b}. While the $e$EDM enhancement coefficient for the Tl atom is about 500, for heavy diatomic molecules, it is much larger, reaching approximately $10^9$ for the ThO molecule. This happens due to the $\Lambda$($\Omega$)-doubling of diatomic molecules with open shells, making them particularly suitable for investigating $\mathcal{P}$- and $\mathcal{T}$,~$\mathcal{P}$-odd effects. The larger is the $\Omega$ value, the smaller is the $\Omega$ splitting and, thus, the stronger are the parity nonconserPhysicalvation effects. The ground state in the YbF molecule possesses $\Omega=1/2$, and in the search for the $e$EDM in the ThO molecule, a metastable state with $\Omega=1$ was used. However, further improvement of the $e$EDM constraints using molecular states with larger $\Omega$ seems to be exhausted. Due to the extremely small splitting for large $\Omega$ in an external electric field, the $\Omega$ sublevels begin to repulse, leading to a saturation of the effect. In Ref.~\cite{Chub14} it was shown that choosing larger values of $\Omega$ only results in diminishing the saturating electric field to such small values that are not convenient for performing experiments. In the current experiments that provide the most stringent constraints on the $e$EDM, either electron spin precession in an external electric field~\cite{ACME18} or trapped molecular ions in a rotating electric field~\cite{JILA23} are investigated. However, in Refs.~\cite{Chub17,Chub23,Chub19-1,Chub19-2,Chub19-3,Chub21,Chekh}, an alternative approach based on the $\mathcal{T}$, $\mathcal{P}$-odd Faraday rotation in high-finesse optical cavities has been proposed and developed, which holds promise for future research.

Theoretical predictions for the $e$EDM value within the Standard Model (SM) are rather uncertain and far from the experimental bounds. The same applies to the $\mathcal{T}$,~$\mathcal{P}$-odd electron-nucleon interaction. It is convenient to express the magnitude of these interactions via the equivalent $e$EDM $d_e^{\text{eqv}}$. This quantity is defined as an $e$EDM that produces the same linear Stark shift in the same electric field as that produced by an electron-nucleon $\mathcal{T}$,~$\mathcal{P}$-odd interaction in a particular atomic system. The first and the only attempt for an accurate calculation of the $e$EDM within the SM was made in Ref.~\cite{Hoog90}. It was found that the $e$EDM can arise when the electron interacts with an external electric field via a three-loop vertex containing one quark loop with 5 quark propagators and 3 exchanges by $W$ bosons. Then, $\mathcal{CP}$ violation occurs due to the phase (an imaginary unity) arising from the Cabibbo-Kobayashi-Maskawa (CKM) quark -- $W$ boson interaction matrix. The twelve-dimensional loop integrals over the particle momenta were evaluated with account for the antisymmetric structure of the CKM matrix with respect to the quark flavors, i.e., the Glashow-Iliopoulos-Maiani (GIM) mechanism. The result obtained in Ref.~\cite{Hoog90} for the $e$EDM was $d_e\sim 10^{-38}$~$e$cm. However, it was later shown in Ref.~\cite{Pos91} that the total three-loop result for the $e$EDM should be exactly zero. This happens because the integrand represents the quark self-energies or vertices which are symmetric with respect to the quark flavors. Combined with the antisymmetry following from the CKM matrix, this gives an exact zero result, i.e., the contributions of different quark flavors cancel each other. As noted in Ref.~\cite{Pos91}, a nonzero result could arise with an addition of one gluon exchange, in a four-loop approximation.  A four-loop result was estimated in Ref.~\cite{Pos14} using only the GIM mechanism. The gluon exchange contribution was approximated as $\alpha_{\text s}/4\pi \approx 1/10$, where $\alpha_{\text s}$ is the strong interaction constant. The result of this estimate was $d_e\approx 10^{-44}$~$e$cm. In Ref.~\cite{Pos14}, a particular model for the $\mathcal{T}$,~$\mathcal{P}$-odd electron-nucleon interaction was suggested, namely, the two-photon exchange with one $\mathcal{C}\mathcal{P}$-violating vertex on the electron line. The estimated value of $d_e^{\text{eqv}}$ reported in Ref.~\cite{Pos14} was $d_e^{\text{eqv}}\approx 10^{-38}$~$e$cm. While the estimate for the $e$EDM in Ref.~\cite{Pos14} was made at the free quark level, the estimate for the $\mathcal{CP}$-violating two-photon exchange between an electron and a nucleus was made at the hadronic level, i.e., for quarks coupled into hadrons. Recently, an accurate semiempirical evaluation of the $e$EDM at the hadronic level was performed in Ref.~\cite{Yam21} with $d_e \approx 10^{-39} ~ e\text{cm}$. Estimates made for $d_e$ in Ref.~\cite{Yam21} at the free quark level gave a result of $d_e \approx 10^{-50} ~ e\text{cm}$ which strongly differs from Ref.~\cite{Pos14}. In a recent paper~\cite{Ema22}, a semi-phenomenological calculation (at the hadronic level) of the $\mathcal{T}$,~$\mathcal{P}$-odd electron–nucleon interaction via the exchange of a neutral $K$ meson was performed and yielded the ``benchmark'' equivalent $e$EDM $\sim 10^{-35}$~$e$cm.

Another model was suggested in Ref.~\cite{Chub16}. It consists of a Higgs-boson exchange between an electron and a nucleus in an atomic system. The interaction of an atomic electron with the Higgs boson is described by the $\mathcal{C}\mathcal{P}$-odd three-loop vertex similar to that employed in Ref.~\cite{Hoog90} to describe the interaction of an electron with an external electric field. The proposed idea was that this $\mathcal{T}$,~$\mathcal{P}$-odd contribution, unlike the corresponding contribution to the $e$EDM, should not vanish in the three-loop approximation, since the ``Higgs charges'' of the quarks, contrary to their electric charges, are flavor dependent. Rough estimates at the free quark level gave the upper bound $d_e^{\text{eqv}}< 10^{-40}$~$e$cm~\cite{Chub16}. Note that the estimate in Ref.~\cite{Chub16} was made by means of a comparison with analogous estimates for the $e$EDM from the previous studies~\cite{Hoog90,Pos14} without applying the GIM mechanism explicitly. Additional motivation for considering this effect is an opportunity to indirectly observe the Higgs boson by atomic physics methods. Here we can recall a somewhat similar situation concerning the $Z$ boson, which is also heavy and observed only due to the fact that its exchange between the electron and nucleus violates $\mathcal{P}$ symmetry. Moreover, it is the atomic experiment that gives the best constraints on a mass of the second $Z$ boson. A Higgs-boson exchange is always present in atomic physics as $H$ interacts with all fermions. However, this interaction is strongly screened by electromagnetic forces since the Higgs-boson exchange itself does not violate any symmetry unlike the $Z$-boson exchange. The possibility of observing the former based on $\mathcal{T}$~and~$\mathcal{P}$ symmetry violations is of current interest to us.

In the present paper, we provide a more accurate estimate of $\mathcal{T}$,~$\mathcal{P}$-odd exchange of the Higgs boson between an electron and a nucleon based on the GIM mechanism. Our paper is organized as follows. In the next Section, we present an instructive proof of the dependence of the electron EDM on its mass by analyzing the three-point $\gamma W^+ W^-$ vertex. In Sec.~\ref{sec:3loop} we explicitly demonstrate that the total three-loop contribution vanishes for the $\mathcal{T}$, $\mathcal{P}$-odd electron-nucleon interaction via the exchange of the Higgs boson. In Sec.~\ref{sec:4loop} we evaluate the suppression factors following from the flavor antisymmetry as consequences of the GIM mechanism at the four-loop level for the considered effect. We provide a final estimate for the $\mathcal{CP}$-odd electron-nucleon interaction coupling constant in the Higgs boson exchange mechanism and evaluate the corresponding equivalent $e$EDM. Besides, we provide the GIM estimate of the $e$EDM at the four-loop level. It proves to differ from the previous predictions~\cite{Pos14,Yam21}.

Throughout the text, we employ the relativistic units $\hbar=c=1$ ($\hbar$ is the Planck constant, $c$ is the speed of light). The charge units correspond to $\alpha = e^2/(4 \pi)\approx 1/137$ ($e>0$ is the elementary charge, $\alpha$ is the fine-structure constant).

\section{Mass dependence of the magnetic and electric dipole moments of the electron}\label{sec:mass}
\subsection{General remarks}\label{subsec:gen_rem}

\begin{figure}[t]
\begin{center}
\includegraphics[width=0.55\columnwidth]{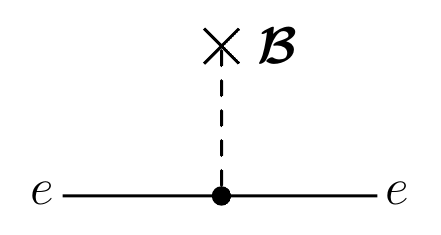}
\caption{\label{fig:magn_mom} Tree-level Feynman diagram for the interaction of an electron with a uniform constant magnetic field $\boldsymbol{\mathcal{B}}$.}
\end{center}
\end{figure}

It is widely known that the magnetic dipole moment of the electron is expressed in terms of the Bohr magneton 
$\mu_{\text B}= e/(2m_e)$ and is inversely proportional to the electron mass $m_e$. This relationship can be derived either directly from the Dirac equation or, e.g., from the tree-level Feynman diagram depicted in Fig.~\ref{fig:magn_mom} (see, e.g., Refs.~\cite{LL4,Akhiez}). 
In Fig.~\ref{fig:magn_mom}, the interaction of electron $e$ with a uniform constant magnetic field $\boldsymbol{\mathcal{B}}$ is depicted.
Moreover, the dependence on the electron mass remains the same even when considering radiative quantum electrodynamics (QED) corrections, which include only virtual electron-positron pairs and photons. Incorporation of other virtual particles (muons, quarks, gauge bosons, etc.) into the theory would introduce contributions with different dependencies on the electron mass since the mass dimensions of different particle types can cancel each other. Nevertheless, these contributions are very small compared to the leading radiative QED corrections.  

\begin{figure}[t]
\begin{center}
\includegraphics[width=0.8\columnwidth]{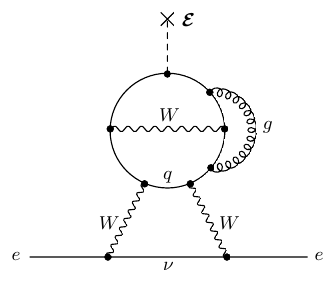}
\caption{\label{fig:edm} Four-loop Feynman diagram for the interaction of the electron EDM with a uniform constant electric field $\boldsymbol{\mathcal{E}}$. The solid lines denote fermions: electron $e$, neutrino $\nu$, and quarks $q$. The wavy lines denote the $W$ bosons, the curly line denotes the gluon $g$, and the dashed line with a cross corresponds to the interaction with an external field.}
\end{center}
\end{figure}

Alternatively, as stated, e.g., in Refs.~\cite{Pos14, Ema22}, the electron electric dipole moment ($e$EDM) which arises only due to the violation of $\mathcal{T}$ ($\mathcal{CP}$) and $\mathcal{P}$ symmetries, should be at least linear in $m_e$. In Refs.~\cite{Pos14, Ema22}, it is explained by the conservation of electron chirality in the SM. However, it is instructive to directly identify the mass dependence of the $e$EDM from Feynman diagrams corresponding to the interaction of an electron with a uniform constant electric field $\mathcal{E}$. Consider such an interaction through the three-point $\gamma W^+ W^-$ vertex. As was mentioned in the Introduction, the first nonvanishing contribution to the $e$EDM appears at the four-loop quark-gluon level (see Fig.~\ref{fig:edm}). 
In Fig.~\ref{fig:edm}, a uniform constant electric field $\boldsymbol{\mathcal{E}}$ should be attached to each quark and $W$ boson propagators. The diagram presented in Fig.~\ref{fig:edm} is irreducible. For such diagrams, it is possible to calculate the energy shift in the following way. If the $e$EDM is nonzero, then the electron in a uniform constant electric field should exhibit an energy shift 
\begin{equation}
 \label{shift}
\Delta E = - \boldsymbol{d}_e \boldsymbol{\mathcal{E}}.
\end{equation}
Then, using the connection between the nondiagonal matrix element of the $S$ matrix $S_{AA'}$ between different electron states $|A\rangle$ and $|A'\rangle$ with energies $E_A$ and $E_{A'}$, respectively, and the amplitude $U_{AA'}$  of the process depicted in Fig.~\ref{fig:edm}
\begin{equation}
 \label{S_matrix}
S_{AA'} = - 2\pi i \delta(E_{A}-E_{A'}) U_{AA'}, 
\end{equation}
one finds
\begin{equation}
 \label{shift2}
 \Delta E = U_{AA} \equiv \langle A | U | A \rangle.
\end{equation}
This rule corresponds to the introduction of the effective potential energy operator~\cite{Akhiez}.

\subsection{Two-loop $\gamma W^+ W^-$ vertex}\label{subsec:two_loop}

\begin{figure}[t]
\begin{center}
\includegraphics[width=0.9\columnwidth]{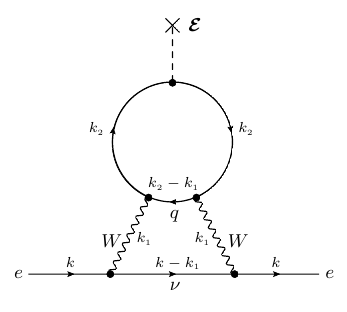}
\caption{\label{fig:2loop} Two-loop Feynman diagram for the interaction of the electron EDM with a uniform constant electric field $\boldsymbol{\mathcal{E}}$ with specified particles' momenta ($k$, $k_1$ and $k_2$). Notations are the same as in Fig.~\ref{fig:edm}.}
\end{center}
\end{figure}

In this Section, we are only interested in the matrix structure of the corresponding amplitude and in the extraction of the electron mass. Therefore, one can neglect the GIM mechanism which gives additional suppression depending on the quark masses. Also, for the sake of simplicity, it is sufficient to consider only a two-loop diagram. Although this contribution does not contain the $\mathcal{CP}$-odd phase, the resulting form of the effective $e$EDM operator and the proportionality between the $e$EDM and the electron mass are independent of the number of $W$ boson and gluon exchange insertions in the quark loop. The generalization of this fact to the many-loop approximation will be separately addressed below. First, let us attach a constant external electric field $\mathcal{E}$ to one of the quark propagators in the loop (see Fig.~\ref{fig:2loop}).
One can express a nondiagonal matrix element of the $S$ matrix for the considered diagram in coordinate representation and perform a trivial integration over all time-space four-vectors. Then, it becomes possible to extract an energy delta-function in accordance with Eq.~\Br{S_matrix}. However, there is an important peculiarity to mention. One common way to treat an external classical constant electric field is to consider standard expressions for the electromagnetic four-potential $\mathcal{A}_{\mu} (q)$ and then, in the final expression, pass to the static limit $q\rightarrow 0$. However, here we will follow another equivalent approach. We incorporate into the Feynman diagram technique a classical four-potential of a constant external electric field as $\mathcal{A}^{\mu} (t) = (0, -\boldsymbol{\mathcal{E}} t)^t$. Then, the integration over the time variable at the quark -- external background vertex gives rise to a derivative of the delta-function. In other words, one of the quark propagators $S^q(p)$ to which an external field is attached, should be replaced by its energy derivative $-\partial S^q(p)/\partial p_0$. For the derivative of the quark propagator, one can use the obvious identity
\begin{equation}
 \label{der_quark_prop}
\frac{\partial S^q(p)}{\partial p_0} = -S^q(p) \gamma_0 S^q(p). 
\end{equation}
In Fig.~\ref{fig:2loop} we indicate the particle momenta. Now we can express the structure of a two-loop amplitude in terms of propagators and simple vertices (omitting spinor indices) as
\begin{widetext}
\begin{eqnarray}
\label{two_loop_amp}
U_{AA}^{\text{two-loop}} \sim \int \! dk_1 dk_2 \overline{\psi}_e(k)\Gamma_{\mu_1}^{e\nu W} S^{\nu}(k-k_1) \Gamma_{\mu_2}^{e\nu W} \psi_e(k) D^{\mu_1\mu_3}(k_1) D^{\mu_4\mu_2}(k_1) \mathcal{E}^i \text{Tr}\left[ \Gamma_{\mu_3}^{e\nu W} \frac{\partial S^q(k_2)}{\partial k_2^0} \Gamma_{i}^{qq\mathcal{A}}  S^q(k_2) \Gamma_{\mu_4}^{e\nu W}   S^q(k_2-k_1)\right].
\end{eqnarray}
\end{widetext}
Here $\psi_e$ denotes the electron wave function in momentum representation, $S^{\nu}$, $D_{\mu\nu}$ and $S^q$ are the neutrino, $W$-boson and quark propagators, respectively. Below, in Sec.~\ref{subsec:prop_vert}, we provide the expressions for these propagators and different vertices representing $W$ \--- lepton and quark \--- external background interactions denoted by $\Gamma_{\mu}$.

\subsection{Propagators and vertices}\label{subsec:prop_vert}

Omitting the spinor indices, we write the quark propagators in the standard form:
\begin{equation}
 \label{quark_prop}
S^q(p)=\frac{\gamma_{\mu}p^{\mu}+m_q}{p^2-m_q^2+i0}.
\end{equation}
Here $\gamma_{\mu}$ are the Dirac matrices. We also define $\gamma_5 = i \gamma_0 \gamma_1 \gamma_2 \gamma_3$. The neutrino propagator reads
\begin{equation}
 \label{neut_prop}
S^{\nu}(p)=\frac{\gamma_{\mu}p^{\mu}}{p^2+i0}.
\end{equation}
The standard expression for the $W$ boson propagator reads
\begin{equation}
 \label{W_prop}
D^W_{\mu\nu}=\frac{-1}{p^2-m_W^2+i0}\left[g_{\mu\nu}+\frac{(\xi-1)p_{\mu}p_{\nu}}{p^2-\xi m_W^2}\right],
\end{equation}
where $\xi$ is the gauge parameter: $\xi=1$ is the Feynman gauge, $\xi=0$ is the Landau gauge, and $\xi=\infty$ corresponds to the unitary gauge~\cite{Cheng84}. Following Refs.~\cite{Shab78,Hoog90}, we choose the unitary gauge since in this gauge it is not necessary to include the interaction with the Faddeev-Popov (FP) scalar ghosts. In the unitary gauge, the FP propagators turn to zero~\cite{Cheng84}. Generally speaking, with the unitary gauge, the individual terms of the amplitude diverge more strongly in the loop integrals. However, the divergence becomes less severe due to the cancelations between the different terms, i.e., due to the GIM mechanism. In the unitary gauge, the $W$ boson propagator reads
\begin{equation}
 \label{W_prop_unit}
D^W_{\mu\nu,\text{unit}}=\frac{-1}{p^2-m_W^2+i0}\left(g_{\mu\nu}-\frac{p_{\mu}p_{\nu}}{m_W^2}\right).
\end{equation}

The lepton -- $W$ vertices~\cite{Khrip91} and the quark -- $W$ vertices~\cite{Cheng84, Jar85} are given by
\begin{eqnarray}
 \label{lW_vert}
\Gamma_{\mu}^{e\nu W} &=& \frac{ig}{2\sqrt{2}} \gamma_{\mu}\left(1-\gamma_5\right), \\
\Gamma_{\mu}^{qq' W} &=& \frac{ig}{2\sqrt{2}} \gamma_{\mu}\left(1-\gamma_{5}\right)V_{qq'},
\label{qW_vert}
\end{eqnarray}
where $V_{qq'}$ are the matrix elements of the CKM matrix.
The quark -- external background vertices read~\cite{Cheng84}
\begin{equation}
 \label{qA_vert}
\Gamma_{\mu}^{qq\mathcal{A}} = ie Q_q \gamma_{\mu},
\end{equation}
where $Q_q$ is the charge of the quark $q$.

A connection between the constant $g$ and the electric charge of an electron is~\cite{Cheng84}
\begin{equation}
 \label{eg}
e=g\sin \theta_W,
\end{equation}
where $\theta_W$ is the Weinberg angle. For the $\theta_W$ angle  we adopt the value
\begin{equation}
 \label{Wein_angle}
\sin^2 \theta_W\approx 0.22.
\end{equation}

In the next Sections, we will also need the scalar Higgs boson propagator
\begin{equation}
 \label{prop_H}
D^H=\frac{1}{p^2-m_H^2+i0}.
\end{equation}
%

\subsection{Extraction of the electron mass}\label{subsec:el_mass}

First, consider the neutrino propagator in \Eq{two_loop_amp}. It is natural to define the $e$EDM according to \Eq{shift} for an electron in the rest frame, so one can set $k^\mu = (m_e,0,0,0)^t$. Omitting the imaginary part, the neutrino propagator has the form
\begin{equation}
 \label{neut_prop2}
S^{\nu}(k-k_1)=\frac{\gamma_\nu (k-k_1)^\nu}{(m_e-k_1^0)^2 -\boldsymbol{k}_1^2}.
\end{equation}
From the expression for the amplitude~\Br{two_loop_amp} corresponding to Fig.~\ref{fig:2loop} it follows that the typical loop momenta contributing to the integral are  $k_1^2 \sim m_W^2$~or~$m_t^2$ (see also a general discussion of divergencies in Sec.~\ref{sec:4loop}), where $m_t$ is the mass of the heaviest $t$ quark. Since $m_W\gg m_e$, one can expand \Eq{neut_prop2} in a series,
\begin{eqnarray}
 \label{neut_prop3}
S^{\nu}(k_1)&=& \frac{\gamma_\nu (k-k_1)^\nu}{k_1^2}\left\{1+ \frac{2k_1^0m_e}{k_1^2} \right. \nonumber \\
&+&   \left. \left[4\left(\frac{k_1^0}{k_1}\right)^2-1
 \right]\frac{m_e^2}{k_1^2}  +     \mathcal{O}\left(  \frac{m_e^3}{m_W^3}\right) \right\}.
\end{eqnarray}
Next, let us consider the numerator of the neutrino propagator $\gamma_\nu (k-k_1)^\nu$ in \Eq{two_loop_amp} and show that the contribution of either the first or second term to the amplitude vanishes depending on the order of expansion of the denominator in \Eq{neut_prop3}. For this purpose, we write the matrix structure of the trace (omitting the denominators):
\begin{multline}
(1+\gamma_5)\gamma_{\mu_3} (\gamma_\nu k_2^\nu + m) \gamma_0 (\gamma_\nu k_2^\nu + m) \\
{}\times \gamma_i (\gamma_\nu k_2^\nu + m) \gamma_{\mu_4}   [\gamma_\nu (k_2-k_1)^\nu + m'].
\label{tr_2loop}
\end{multline}
Since only the trace of even number of $\gamma$ matrices is nonzero, one should take into account an even number of $\gamma_\nu k^\nu$ combinations from the propagators. The denominators of the lepton propagators have four-momenta squared, then it is evident that changing the sign of $k_1$ simultaneously with $k_2$ does not lead to the change of the trace. The $W$ boson propagators in~\Eq{two_loop_amp} are also even in four-momentum $k_1$. 

Then, it turns out that the second term of the neutrino propagator with $\gamma_\nu k_1^\nu$ leads to the odd in $k_1$ contribution to the integrand for odd terms in curly braces of \Eq{neut_prop3}. The remaining term of the neutrino propagator with $\gamma_\nu k^\nu$ leads to the odd in $k_1$ contribution to the integrand for even terms in braces of \Eq{neut_prop3}.  For such cases, the integral over $k_1$ vanishes which proves the above-mentioned conclusion. 

Accordingly, the expansion of the neutrino propagator which leads to a nonzero contribution has the following form:
\begin{eqnarray}
 \label{neut_prop4}
S^{\nu}(k_1)&=& \frac{m_e}{k_1^2} \left[\gamma_0 - \frac{2\gamma_\mu k_1^\mu k_1^0}{k_1^2} +\mathcal{O}\left( \frac{m_e^2}{m_W^2}\right)\right].
\end{eqnarray}
From \Eq{neut_prop4} it follows that this is a good approximation that the $e$EDM is at least linear in an electron mass. In fact, the $e$EDM can be presented as
\begin{equation}
 \label{d_m}
d_e \sim m_e \left[1+ \mathcal{O}\left( \frac{m_e^2}{m_W^2}\right)\right].
\end{equation}
In this connection, it is also interesting to see where the effective $e$EDM operator arises. It can be shown by the following transformation of the matrix structure of \Eq{two_loop_amp}:
\begin{widetext}
\begin{eqnarray}
\label{two_loop_amp_matrix}
U_{AA}^{\text{two-loop}} \sim (1-\gamma_5) \left(g_{\mu_10} g_{\mu_2\sigma} +g_{0\mu_2}g_{\mu_1\sigma}- g_{\mu_1\mu_2}g_{0\sigma}-i\epsilon_{\sigma \mu_1 0\mu_2} \right) \gamma^{\sigma}  D^{\mu_1\mu_3}(k_1) D^{\mu_4\mu_2}(k_1) \mathcal{E}^i  T_{\mu_3\mu_4 i} ,
\end{eqnarray}
\end{widetext}
 where $\epsilon_{\mu \nu \lambda \sigma}$ is the Levi-Civita tensor and $T_{\mu_3\mu_4 i}$ is the structure of the trace. It is known that only the pseudoscalar part contributes to the effective $e$EDM operator~\cite{Khrip91}. The expression~\Br{two_loop_amp_matrix} contains two vectors $\gamma^{\sigma}$ and $\mathcal{E}^i$, so it is obvious that after all integrations over momenta the only pseudoscalar quantity that can be constructed is $\gamma_5 \boldsymbol{\gamma}\boldsymbol{\mathcal{E}}$. It coincides with the $e$EDM operator~\cite{Khrip91}.

 Generalization for the many-loop approximation can be done via Euler's formula for the planar graphs. Any insertion of the boson propagator into the loop adds two vertices ($v$) and one loop ($l$). Since $v-p+l=1$ ($p$ is the number of propagators), it adds three propagators (two fermionic ones in the trace and one external of the boson type). A moment's consideration shows that all the manipulations in this subsection are valid for the general case since the proof is formally based only on the symmetrical properties. Attaching an external field to the $W$ boson propagator does not change the situation. Actually, in this case one has also an even number of vertices inside a quark loop, which leads to considering an even number of $\gamma_\nu k^\nu$ combinations from the propagators.

 In the end of this Section, we will give a couple of notes. The first is as follows. Here we define the EDM for an electron in the rest frame. However, similar derivations can be made for electrons with energy $E\ll m_W$. Then, the operator $\gamma_\nu k^\nu$ being sandwiched between Dirac bispinors will immediately yield the leading linear order for the EDM with respect to the electron mass. The second note concerns the scalar-pseudoscalar $\mathcal{T}$, $\mathcal{P}$-odd interaction between an electron and a nucleon (see Secs.~\ref{sec:3loop}~and~\ref{sec:4loop}). Applying the same ideas and taking into account an even number of $\gamma$ matrices at the vertices within the quark loop (the scalar boson-quark vertex does not contain a $\gamma$ matrix), we conclude that the $\mathcal{T}$,~$\mathcal{P}$-odd coupling constant is also at least linear in the electron mass.

\section{Three-loop contribution of the $\mathcal{T}$, $\mathcal{P}$-odd electron-nucleon interaction as exchange by Higgs boson}\label{sec:3loop}

An amplitude of the scalar-pseudoscalar interaction between an electron and a nucleon in atomic systems was introduced in Refs.~\cite{San75,Gor79} as an interaction of the electron pseudoscalar and nucleon scalar currents,
\begin{equation}
\label{CSP}
U_{eN}=  \frac{G_\text{F}}{\sqrt{2}} C_{\text{SP}} \bar{e} i \gamma_5 e\bar{N}N,
\end{equation}
where $C_{\text{SP}}$ is the dimensionless $\mathcal{CP}$-odd electron-nucleon interaction coupling constant that depends on the particular model and $G_\text{F}$ is the Fermi constant. While the model was not specified, it was assumed that the interaction of the atomic electron with all the nucleons in the atomic nucleus is coherently enhanced, so that the interaction of the electron with the nucleus can be obtained by multiplying~\Eq{CSP} by the atomic number (number of nucleons) $A$. Later, in Ref.~\cite{Khrip91}, a general statement was formulated that the Hermitian $\mathcal{T}$, $\mathcal{P}$-odd interaction between an electron and a nucleus in an atomic system can be presented only as a scalar-pseudoscalar interaction. However, this statement concerns only the lowest-order approximations. Another possibility is a tensor-pseudotensor interaction that is related to the $e$EDM effect. In this case, the corresponding amplitude can be presented as a convolution of the electron pseudotensor $\gamma_5 \sigma_{\mu\nu}$ [$\sigma_{\mu\nu}=1/2 (\gamma_\mu\gamma_\nu-\gamma_\nu\gamma_\mu)$] with a tensor of an external electromagnetic field $F^{\mu\nu}$:
\begin{equation}
\label{tens_edm}
U_{e}=  \frac{d_e}{2}  \bar{e} i \gamma_5 \sigma_{\mu\nu}F^{\mu\nu} e.
\end{equation}

As was mentioned in the Introduction, the $e$EDM vanishes in the three-loop approximation~\cite{Pos91}. The proof of this statement is based on the symmetry of this contribution with respect to the quark flavors combined within the GIM-antisymmetrization mechanism.  Additionally, the equality of electric charges among all the up-type quarks ($u,c,t$) and among all the down-type quarks ($d,s,b$) was crucial for obtaining a zero result. This observation inspired us to propose another mechanism for generating the $\mathcal{T}$, $\mathcal{P}$-odd effects, specifically concerning the electron-nucleon interaction. The interaction described in~\Eq{CSP} resembles an exchange mediated by a heavy scalar particle with $\mathcal{C}\mathcal{P}$ violation. Therefore, it is reasonable to assume that this exchange is associated with the Higgs boson~\cite{Chub16}. In this context, the coupling vertex between the Higgs boson and quarks, often referred to as the ``Higgs charge'' of the quark, is proportional to the quark mass. Since all of the quark flavors possess different masses, it was initially believed that this should prevent cancelation and lead to a nonzero contribution within the three-loop approximation~\cite{Chub16}. However, in this Section, we will explicitly demonstrate that, despite this expectation, the total three-loop contribution, in fact, vanishes for the Higgs-exchange mechanism.

The unique source of $\mathcal{CP}$($\mathcal{T}$) violation in the SM is the complex phase in the CKM matrix. This phase becomes nontrivial, i.e., it cannot be eliminated by any quark rotations, when three generations of quarks are considered. Consequently, the $\mathcal{CP}$-odd flavor structure of the quark loop can be expressed as~\cite{Pos91}
\begin{eqnarray}
\label{loop_struct}
  U_a &\sim& 2i J [t(dus-sud+bud-dub-bus+sub) \nonumber\\
  &+& u(dcs-scd+scb-bcs+bcd-dcb) \nonumber\\
  &+& c(dts-std+stb-bts+btd-dtb)].
 \end{eqnarray}
Here we use the standard parametrization for the Jarlskog invariant $J$~\cite{Jar85} which is given by the product of four CKM matrix elements $V_{ij}$. It amounts to~\cite{Tan18}
\begin{equation}
    \label{Jar}
    J = \text{Im} \, (V_{us} V_{td} V_{ud}^* V_{ts}^*) \approx 3.2\times 10^{-5}.
\end{equation}
The letters $u, d,c, s, t,b$ in Eq.~\eqref{loop_struct} denote the propagators of the corresponding quarks. Only the imaginary part of the product of $V_{ij}$ in \Eq{Jar} contributes to the $\mathcal{CP}$-odd effects. In~\Eq{loop_struct}, each product of the quark propagators allows for cyclic permutations such as $tdus=dust=ustd=stdu$. Thus, at the pure quark level, any such expression should be represented by a Feynman diagram containing no fewer than four ``quark -- $W$ boson -- quark'' vertices.

\begin{widetext}

\begin{figure}[H]
\begin{center}
\begin{minipage}[h]{0.31\linewidth}
\center{\includegraphics[width=1\linewidth]{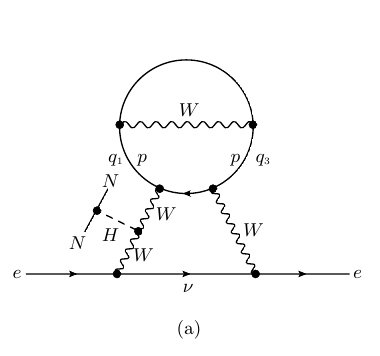}}  \\
\end{minipage}
\hfill
\begin{minipage}[h]{0.31\linewidth}
\center{\includegraphics[width=1\linewidth]{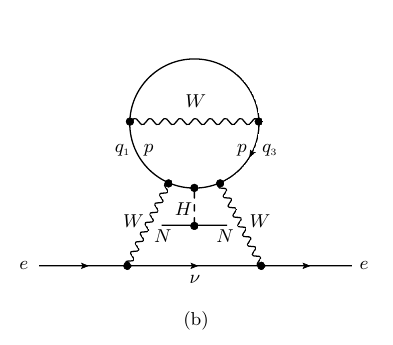}} \\
\end{minipage}
\hfill
\begin{minipage}[h]{0.31\linewidth}
\center{\includegraphics[width=1\linewidth]{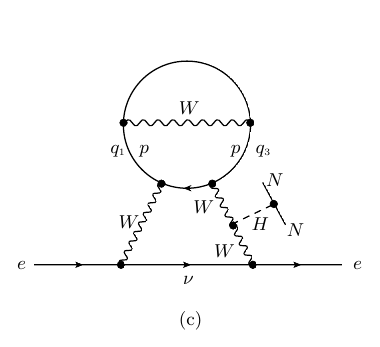}} \\
\end{minipage}
\vfill
\begin{minipage}[h]{0.31\linewidth}
\center{\includegraphics[width=1\linewidth]{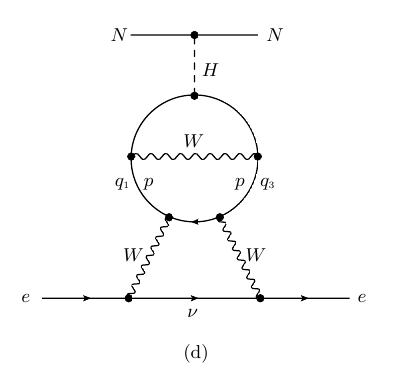}} \\
\end{minipage}
~~~~~
\begin{minipage}[h]{0.31\linewidth}
\center{\includegraphics[width=1\linewidth]{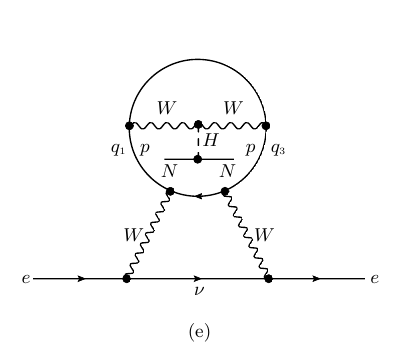}} \\
\end{minipage}
\vfill
\begin{minipage}[h]{0.31\linewidth}
\center{\includegraphics[width=1\linewidth]{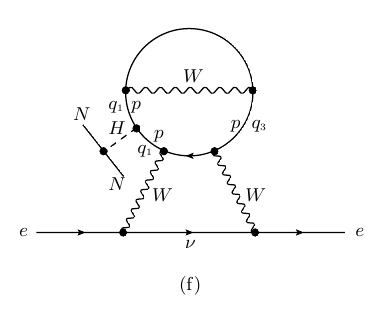}} \\
\end{minipage}
~~~~~
\begin{minipage}[h]{0.31\linewidth}
\center{\includegraphics[width=1\linewidth]{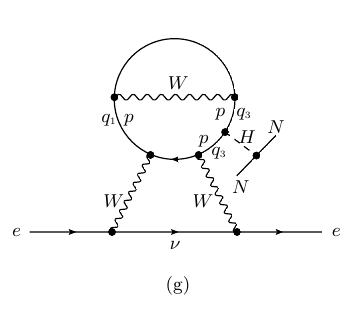}} \\
\end{minipage}
\end{center}
\caption{\label{fig:3loop} Feynman diagrams describing three-loop contribution to the electron-nucleon interaction induced by the Higgs boson exchange. Here $N$ stands for the nucleon and $H$ denotes the Higgs boson. All other notations are the same as in Fig.~\ref{fig:edm}.}
\end{figure}

\end{widetext}

In other words, the $\mathcal{T}$, $\mathcal{P}$-odd Higgs boson exchange mechanism at the three-loop level can be represented as a sum of Feynman diagrams shown in Fig.~\ref{fig:3loop}. These diagrams differ from each other in the arrangement of the quark flavors in a closed fermion loop and in the position of the Higgs boson attachment. In Fig.~\ref{fig:3loop}, we have also set the momentum transferred by the Higgs boson to $q = 0$ since we can replace the Higgs-boson exchange with a four-fermion interaction (see also Sec.~\ref{sec:4loop}). In particular, the expression~(\ref{loop_struct}) indicates that the Feynman diagrams in Fig.~\ref{fig:3loop} should actually be antisymmetrized over the first (left, labeled ``$q_1$'') and third (right, labeled ``$q_3$'') quark propagators with masses $m_1$ and $m_3$, respectively. The upper block is the mass operator $\hat\Sigma$ or the mass operator with an external scalar Higgs boson -- quark or Higgs boson -- $W$ boson vertex $\hat \Sigma^{v+}$. In the unitary gauge and within the vector-minus-axial theory of weak interactions, the structure of the mass operator $\hat\Sigma$ can be represented as~\cite{Shab78}
\begin{equation}
    \label{mass_op}
   \hat\Sigma (p) = \cancel{p} (1-\gamma_5) f(p^2),
\end{equation}
where $ \cancel{p} \equiv \gamma_\mu p^\mu$, $p^\mu$ is the input and output four-momentum, and $f(p^2)$ is some function that depends on $p^2 = \cancel{p}\cancel{p}$ but does not depend on $m_1$ and $m_3$. Note that in this paper, we use a different sign in the definition of the $\gamma_5$ matrix compared to Ref.~\cite{Shab78}. The renormalized mass operator $\hat\Sigma^{(\text{R})}$ has the following form~\cite{Shab78}:
\begin{eqnarray}
    \label{mass_op_r}
   \hat\Sigma^{(\text{R})}  &=& \cancel{p} (1-\gamma_5) \left[f(p^2)-\frac{m_1^2 f_1-m_3^2 f_3}{m_1^2-m_3^2}\right] \nonumber \\
   &-& f_{13}\left[\cancel{p}(1+\gamma_5)-m_1(1+\gamma_5)-m_3 (1-\gamma_5)     \right],
\end{eqnarray}
where
\begin{equation}
f_{13}\equiv\frac{m_1m_3(f_1-f_3)}{m_1^2-m_3^2}
\end{equation}
and $f_i\equiv f(p^2=m_i^2)$.

The analysis carried out in Ref.~\cite{Chub16} demonstrated that only two Feynman diagrams [Figs.~\ref{fig:3loop}(f) and (g)] can contribute to the $\mathcal{T}$, $\mathcal{P}$-odd effect under consideration. In fact, the Feynman diagrams depicted in~Figs.~\ref{fig:3loop}(a)-(e) have the following structure:
\begin{equation}
    \label{ae}
  (1-\gamma_5)  S^{q_1} (p) \hat A (p) S^{q_3} (p) (1+\gamma_5),
\end{equation}
where $\hat A (p)=\hat \Sigma (p)$ for diagrams (a)-(c) and $\hat A (p)=\hat \Sigma^{v+} (p)$ for the sum of (d) and (e). The projection operators $1\pm \gamma_5$ in \Eq{ae} arise from the $W$-boson vertices. Now it turns out that both $S^{q_1}$ and $S^{q_3}$ are sandwiched between these projection operators, resulting in the cancelation of mass insertions in their numerators. Thus, we have
\begin{equation}
    \label{ae2}
  (1-\gamma_5)  \frac{\cancel{p} }{p^2-m_1^2} \hat A (p)  \frac{\cancel{p} }{p^2-m_3^2} (1+\gamma_5).
\end{equation}
It is now obvious that these contributions are symmetric under the interchange of $m_1 \leftrightarrow m_3$ and vanish upon applying the GIM antisymmetrization. This cancelation of contributions for each diagram is independent of the renormalization of its upper block.

We will now demonstrate that the combined terms corresponding to the diagrams in Figs.~\ref{fig:3loop}(f) and (g) also do not contribute to the effect. This sum reads
\begin{equation}
    \label{fg}
  (1-\gamma_5) \left(  S^{q_1}   m_1 S^{q_1} \hat \Sigma  S^{q_3}+  S^{q_1}  \hat \Sigma S^{q_3}m_3 S^{q_3} \right)(1+\gamma_5),
\end{equation}
where the factors $m_1$ and $m_3$ originate from the quark-Higgs-boson vertex. Due to the presence of $\gamma_5$ matrices, one can rewrite \Eq{fg} as 
\begin{equation}
    \label{fg2}
  4(1-\gamma_5)\left[   \frac{ m_1^2\cancel{p}  \hat \Sigma \cancel{p}}{(p^2-m_1^2)^2 (p^2-m_3^2)}  + \frac{ m_3^2\cancel{p}  \hat \Sigma \cancel{p}}{(p^2-m_1^2) (p^2-m_3^2)^2} \right].
\end{equation}
The symmetry of the expression~(\ref{fg2}) under the interchange $m_1 \leftrightarrow m_3$ along with the GIM mechanism leads to a zero result. Establishing symmetry for the renormalized mass operator~\eqref{mass_op_r} requires slightly more intricate evaluations. 
 However, it can be easily shown that~\Eq{fg}, with the first line of the expression for $\hat\Sigma^{(\text{R})}$~\Br{mass_op_r}, is evidently symmetric under the interchange $m_1 \leftrightarrow m_3$. For the square brackets of the second line of \Eq{mass_op_r}, the relation~(\ref{fg}) yields
\begin{equation}
    \label{fg3}
  8(1-\gamma_5)\cancel{p} \left[ \frac{ m_1^3 m_3}{(p^2-m_1^2)^2 (p^2-m_3^2)}  + \frac{ m_3^3 m_1}{(p^2-m_1^2) (p^2-m_3^2)^2} \right]
\end{equation}
and obviously has the same symmetry. Thus, the total three-loop contribution vanishes for the Higgs-exchange mechanism, and the additional gluon loop should be introduced.

\section{Estimation of the four-loop $\mathcal{T}$, $\mathcal{P}$-odd effects within the GIM mechanism}\label{sec:4loop}

\begin{figure}[t]
\begin{center}
\includegraphics[width=0.9\columnwidth]{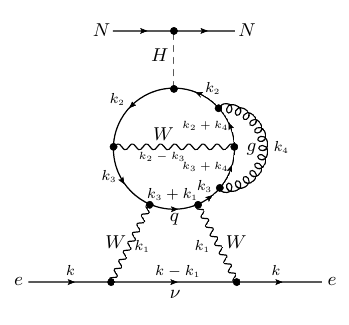}
\caption{\label{fig:4loop} Four-loop Feynman diagram for the $\mathcal{T}$, $\mathcal{P}$-odd electron-nucleon interaction mediated by the Higgs boson exchange with indicated particles' momenta. Notations are the same as in Figs.~\ref{fig:edm}~and~\ref{fig:3loop}.}
\end{center}
\end{figure}

\begin{figure}[t]
\begin{center}
\includegraphics[width=0.95\columnwidth]{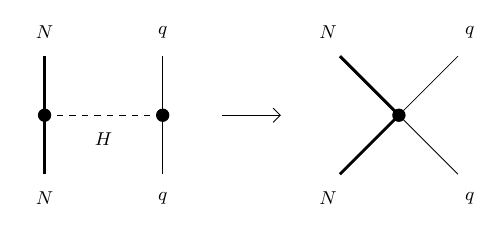}
\caption{\label{fig:Higgs_4ferm} The interaction between the nucleon $N$ and quark $q$ via the Higgs-boson exchange is replaced with a four-fermion interaction vertex. Notations are the same as in Fig.~\ref{fig:3loop}.}
\end{center}
\end{figure}

In this Section, let us provide an estimate of the $\mathcal{T}$, $\mathcal{P}$-odd electron-nucleon interaction via the Higgs-boson exchange at the four-loop level within the GIM mechanism framework. We present a detailed picture of the four-loop diagram, including additional gluon exchange and particle momenta indicated in Fig.~\ref{fig:4loop}.
A special treatment is required for the interaction between the nucleon and the quark loop mediated by the Higgs-boson exchange. We replace this interaction with the four-fermion vertex according to Fig.~\ref{fig:Higgs_4ferm}. This corresponds to the following approximation:
\begin{eqnarray}
 \label{4ferm}
\langle Nq|\Gamma^{\mu} D^H_{\mu\nu} \Gamma^{\nu} | Nq\rangle &\approx& \frac{H_NH_q}{m_H^2} \langle N|N\rangle\langle q|q\rangle \nonumber \\
&=& \frac{G_{\text{F}}}{\sqrt{2}} \frac{\sqrt{2}H_NH_q}{G_\text{F} m_H^2},
\end{eqnarray}
where $m_H$ is the Higgs boson mass, $H_N$ is the Higgs charge of the nucleon, and $H_q$ is the Higgs charge of the quark $q$. In \Eq{4ferm} we isolate the factor $G_{\text{F}}/\sqrt{2}$ according to the definition of the $C_{\text{SP}}$ constant in \Eq{CSP}.

The Higgs charge of the quark with the mass $m_q$ is defined as~\cite{Ok82}
\begin{equation}
 \label{quark_H}
 H_q= \frac{m_q}{\eta},
\end{equation}
where $\eta$ is the vacuum expectation value of the Higgs field~\cite{Ok82},
\begin{equation}
 \label{eta}
 \eta\approx 246 \; \text{GeV}.
\end{equation}
Due to the smallness of the Higgs charge of the electron
\begin{equation}
 \label{e_H}
  H_e= \frac{m_e}{\eta}\sim 10^{-5},
\end{equation}
we consider the $\mathcal{C}\mathcal{P}$-violating vertex involving a quark loop connected to the electron, rather than to the nucleon line. The Higgs charge of the nucleon has a nontrivial origin~\cite{Ok82}: the main contribution comes from the interaction with the gluons (not with the quarks) inside the nucleon. This interaction leads to the conclusion that the Higgs charge is approximately the same for both the proton and neutron, resulting in a common value~\cite{Ok82}
\begin{equation}
 \label{N_H}
  H_N\approx-0.3 \times 10^{-3}.
\end{equation}
It is important to emphasize that a direct gluon exchange between the nucleon and the quark loop is not feasible since the nucleon is a colorless particle. Note one more time that, unlike the case of the $e$EDM interaction with an external electric field, in the case of the electron-nucleon interaction via the Higgs-boson exchange, the amplitude does not depend on the electric charges of quarks, but depends on their Higgs charges. This fact will have advantages for our following estimates.

To analyze the divergences arising from the individual terms in the amplitude to the loop integral within the unitary gauge, we can adopt simplified asymptotic expressions for the particles' propagators. In this approach, we will neglect their matrix (and color) structure, focusing instead on their momentum dependence. The propagators can be approximated as follows:
\begin{equation}
 \label{asymp_q}
 \lim_{p\to\infty} S^{q}(p)=\frac{1}{p}
\end{equation}
for the quarks,
\begin{equation}
 \label{asymp_w}
     \lim_{p\to\infty} D^{W}(p)=\frac{1}{m_W^2}
\end{equation}
for the $W$ boson,
\begin{equation}
 \label{asymp_g}
     \lim_{p\to\infty} D^{g}(p)=\frac{1}{p^2}
\end{equation}
for the gluon, and
\begin{equation}
 \label{asymp_n}
 \lim_{p\to\infty} S^{\nu}(p)=\frac{m_e}{p^2}
\end{equation}
for the neutrino. The asymptotic behavior of the neutrino propagator~(\ref{asymp_n}) needs an additional remark. In Sec.~\ref{subsec:el_mass} it was shown that for the $e$EDM effect, the neutrino propagator reduces to \Eq{neut_prop4}. Of course, the same holds for the $\mathcal{T}$, $\mathcal{P}$-odd pseudoscalar-scalar electron-nucleon interaction, which leads to the asymptotic expression~(\ref{asymp_n}). Therefore, it follows that the coefficient $C_{\text{SP}}$ should also be at least proportional to the electron mass $m_e$.  
 
Going from $\int d^4k$ to $\int k^3 dk$, the loop integral from Fig.~\ref{fig:4loop} in the asymptotic regime behaves as follows:
\begin{equation}
 \label{loop_as}
 I \sim \frac{m_e}{m_W^6} \int  \frac{k_1 dk_1k_2dk_2k_3dk_3k_4dk_4}{(k_2+k_4)(k_3+k_4)(k_1+k_3)}.
\end{equation}
 The expression for the integrand in \Eq{loop_as} should also include the numerator, which, however, does not influence the asymptotic behavior of the integrand and is therefore neglected. Additionally, we have omitted the angular integration as it does not affect the estimate. From \Eq{loop_as}, one can observe significant ultraviolet divergences (up to linear order) for some momenta. This is a consequence of the unitary gauge, indicating that the typical loop momenta circulating in the loop are very large [$k\sim m_W (m_t)$]. Now it is straightforward to estimate the integral by introducing a cutoff energy $\Lambda$. For all terms in \Eq{loop_as}, the integral behaves as
\begin{equation}
 \label{loop_cut}
 I \sim  \left(\frac{m_e}{m_W}\right)\left(\frac{\Lambda}{m_W}\right)^5.
\end{equation}
We observe that the loop integral itself (excluding factors from the vertices) is dimensionless. However, due to the strong divergencies, the order of magnitude of the loop integral $I$ should be fully defined by the GIM mechanism. In fact, the beneficial influence of the GIM mechanism is to reduce these divergencies (at least to the logarithmic ones). For the GIM estimate, we can set all momenta $k_i=\Lambda > m_t$, and after eliminating divergences we take $\Lambda \rightarrow m_t$.

All of the terms in square brackets in \Eq{loop_struct} differ only by the quark masses $m_q$ in the propagators. If we neglect all the quark masses compared to the quark momenta in the propagators, the result will be exactly zero. A nonzero result arises only if we expand the propagators in terms of $\left(m_q/\Lambda\right)^2$ and take into account the corrections.  Additionally, for the GIM estimate, we can neglect the mass insertions in the propagator numerators (since $m_q \ll k=\Lambda$), implying that linear corrections of the form $m_q/\Lambda$ will not contribute.

The next question is: to which quark propagator should the Higgs-boson vertex be attached? The idea of the work~\cite{Chub16} was to fix the quark $q=t$. However, here we argue that the main contribution arises from the attachment to the $s$ quark. Due to the symmetry properties of the amplitude and four flavor changes, only four of the heaviest quarks ($c,s,t,b$) will contribute significantly to the order-of-magnitude estimate. Since the expansion of the quark propagators and applying the GIM mechanism yield factors that are at least quadratic in the quark masses, while the Higgs boson-quark vertex contributes only linearly in mass, the effect is predominantly governed by the $sHs$ vertex (this is justified below). Here $H$ is the interaction with the Higgs boson, so $sHs \sim s H_s s$ [see \Eq{4ferm}].

Next, by distributing the other quarks in all possible configurations and utilizing the CKM matrix, we obtain the following flavor structure for the amplitude (Fig.~\ref{fig:4loop}):
\begin{multline}
U_a \sim 2iJ \left( sHss\right)(uudc-ccdu+ccbu-uubc+ccdt-ttdc \\
+ ttbc-ccbt+ttdu-uudt+uubt-ttbu). \label{gim_loop_st}
\end{multline}
Recall that if we set all momenta $k=\Lambda$, we can rewrite \Eq{gim_loop_st} in the following form:
\begin{eqnarray}
\label{gim_loop_st2}
U_a \sim 2iJ \left( sHss\right)(d-b)(uuc-ccu+cct-ttc + ttu-uut).
\end{eqnarray}
Then we expand the denominators of \Eq{gim_loop_st2} in terms of $m_q^2/\Lambda^2$ assuming $\Lambda \gg m_t$ and retaining only the leading-order terms, taking into account the inequalities $m_t \gg m_c \gg m_u$ and $m_b \gg m_d$. This leads to the additional dimensionless GIM factor
\begin{eqnarray}
\label{gim_loop_st3}
U_a \sim \frac{m_c^2 m_b^2 m_t^4}{\Lambda^8}.
\end{eqnarray}
Analyzing \Eq{gim_loop_st3}, one concludes that it is justified to attach the Higgs boson vertex to the $s$ quark. Additionally, one should assign the factor $\alpha/( \pi\sin^2\theta_{\text{W}})$ for each $W$ boson-quark loop, where $\alpha/\sin^2\theta_{\text{W}}\sim g^2$ arises from the two $W$ boson vertices [see Eqs.~(\ref{lW_vert}),~(\ref{qW_vert}) and (\ref{eg})] and $\pi$ is a standard loop factor. For the quark-gluon loop, we use the estimate $(\alpha_s/\pi) C_F \sim \alpha_s/\pi  $, where $\alpha_s $ is the strong coupling constant, which is approximately equal to 1 for large momenta $k\sim m_t$ circulating in the loop, and the factor $C_F =4/3$ arises from the summation over colours.

Collecting all the factors and substituting $\Lambda$ by $m_t$, we obtain a final parametric GIM estimate of the $\mathcal{T}$, $\mathcal{P}$-odd electron-nucleon interaction via the Higgs-boson exchange
\begin{equation}
\label{CSP_GIM}
C_{\text{SP}} \sim J \frac{H_NH_s}{G_F m_H^2}\frac{m_e m_c^2m_b^2 m_t}{m_W^6}\frac{\alpha^3 \alpha_s}{\pi^4\sin^6\theta_W}\sim 10^{-28}.
\end{equation}
Here we ignore the additional numerical suppression as well as the possible logarithmic enhancement, which can enlarge this value by a couple of orders of magnitude. This value appears to be much smaller than previously predicted~\cite{Chub16}. Finally, according to the idea of Ref.~\cite{Pos14}, we express the equivalent $e$EDM associated with the considered $\mathcal{T}$, $\mathcal{P}$-odd effect as 
\begin{equation}
\label{equiv_de}
d_e^{\text{equiv}} \sim r C_{\text{SP}} \sim 10^{-48} \,\, e\text{cm},
\end{equation}
where the coefficient $r$ is approximately the same for the most recently announced atomic and molecular systems for the search of $\mathcal{T}$,~$\mathcal{P}$-odd effects ($r\sim 10^{-20}$~$e$cm~\cite{Pos14}).

Finally, let us provide the GIM estimate of the $e$EDM effect at the four-loop level. The primary advantage of the GIM mechanism for the Higgs-boson exchange is that we were able to fix the $s$ quark and subsequently distribute all of the other quarks along the loop. This was possible since the vertex linearly depends on the quark mass. However, for the quark -- external background vertex [\Eq{qA_vert}], the situation is different as it is proportional to the quark charge. Therefore, we should sum up the analogue of \Eq{gim_loop_st2} over the down-type quarks. Recall also that according to \Eq{der_quark_prop}, $q A q \sim qq Q_q q $. Abbreviating $\mathcal{F} (u,c,t)= uuc-ccu+cct-ttc + ttu-uut$ in \Eq{gim_loop_st2}, the flavor structure of the $e$EDM amplitude at the four-loop level reads
\begin{eqnarray}
\label{gim_edm}
U_a^{e\text{EDM}} \sim iJ Q_s\left[s^4(d-b)+b^4(s-d)+d^4(b-s) \right]\mathcal{F} (u,c,t),
\end{eqnarray}
where $q^4\equiv qqqq$ and $Q_s=-1/3$. The trivial expansion of \Eq{gim_edm} yields the following dimensionless GIM factor:
\begin{eqnarray}
\label{gim_edm2}
U_a^{e\text{EDM}} \sim \frac{m_c^2 m_s^2 m_b^4 m_t^4}{\Lambda^{12}}.
\end{eqnarray}
In fact, one should also consider attaching the external electric field to the up-type quarks (with a different coefficient $Q_t = 2/3$). However, this will obviously yield the same result as shown in~\Eq{gim_edm2}. This is due to the symmetry present in~\Eq{gim_edm2}: the masses of the two heaviest up- and down-type quarks ($t$ and $b$) appear in the fourth power, while the other two ($c$ and $s$) appear in the second power. By comparing Eqs.~(\ref{gim_loop_st3}) and (\ref{gim_edm2}), one finds that the GIM mechanism for the $e$EDM leads to an additional suppression factor $\sim m_s^2 m_b^2/\Lambda^4$. Thus, the final parametric estimate of the electron EDM reads
\begin{equation}
\label{edm_gim}
 d_e\sim e J \frac{m_e m_s^2 m_c^2m_b^4}{m_W^6 m_t^4}\frac{\alpha^3 \alpha_s}{\pi^4\sin^6\theta_W}\sim 10^{-48}  \,\, e\text{cm}.
\end{equation}
Note that this GIM estimate of the $e$EDM differs from the previous discussions in Refs.~\cite{Pos14} and~\cite{Yam21}. It incorporates an additional mass suppression factor $m_b^4/m_t^4$ compared to Ref.~\cite{Pos14} and  $m_b^2/m_t^2$ compared to Ref.~\cite{Yam21}. Both of the effects --- the $\mathcal{T}$, $\mathcal{P}$-odd electron-nucleon interaction via the Higgs-boson exchange and the electron EDM itself at the four-loop quark-gluon level --- appear to be very small, on the level of $\sim 10^{-48} \,\, e\text{cm}$.

\section{Conclusion}

In this paper, we investigated the mechanism of the electron pseudoscalar -- nucleon scalar $\mathcal{T}$,~$\mathcal{P}$-odd interaction within the Standard Model as an exchange mediated by the Higgs boson at the quark-gluon level. By analyzing the three-point $\gamma W^+ W^-$ vertex, we showed that the $e$EDM $d_e$ as well as the $\mathcal{CP}$-odd electron-nucleon coupling constant $C_{\text{SP}}$ are at least linear in the electron mass. Moreover, these quantities can be expanded into a series with respect to the small parameter $m_e/m_W$ and contain only its odd powers. Besides, we explicitly demonstrated that the total three-loop contribution vanishes for the Higgs-boson exchange mechanism. A nonzero effect appears at the four-loop quark-gluon level with the insertion of an additional gluon line between the quarks. The estimate of the $\mathcal{CP}$-odd electron-nucleon coupling constant after the reduction of divergencies of the quark loop integrals according to the GIM mechanism yielded the value $C_{\text{SP}} \sim 10^{-28}$. The final result, expressed in terms of the equivalent $e$EDM, is $10^{-48}~e\text{cm}$. This value turns out to be considerably smaller than that predicted and roughly estimated in Ref.~\cite{Chub16}. Finally, we performed the GIM estimate of the $e$EDM at the four-loop quark-gluon level and its order of magnitude coincides with the equivalent $e$EDM obtained for the Higgs-boson exchange mechanism. Our parametric estimate of the $e$EDM differs from the previous studies~\cite{Pos14}~and~\cite{Yam21} by a factor that depends on the mass of the $b$ quark.

\begin{acknowledgments}
This work was supported by the Russian Science Foundation (Grant No. 24-72-10060). The authors would like to thank Prof.~Dr.~Leonti~N.~Labzowsky and Prof.~Dr.~Ivan~S.~Terekhov for helpful discussions.
\end{acknowledgments}


\begin{thebibliography}{99}
%
\bibitem{Pur50} E.~M.~Purcell and N.~F.~Ramsey, Phys. Rev. \textbf{78}, 807 (1950).
%
\bibitem{Chris64} J.~H.~Christenson, J.~W.~Chronin, V.~L.~Fitch, and R. Turlay, Phys. Rev. Lett. \textbf{13}, 138 (1964).
%
\bibitem{BABAR01} B.~Aubert et al. (BABAR Collaboration), Phys. Rev. Lett. \textbf{87}, 091801 (2001).
%
\bibitem{Belle01} K.~Abe et al. (Belle Collaboration), Phys. Rev. Lett. \textbf{87}, 091802 (2001).
%
\bibitem{BABAR04} B.~Aubert et al. (BABAR Collaboration), Phys. Rev. Lett.
\textbf{93}, 131801 (2004).
%
\bibitem{Belle04} Y.~Chao et al. (Belle Collaboration), Phys. Rev. Lett. \textbf{93}, 191802 (2004).
%
\bibitem{Lees12} J~ P.~Lees et. al. (The BABAR Collaboration),
Phys. Rev. Lett. \textbf{109}, 211801 (2012).
%
\bibitem{Zal22} T.~Zalialiutdinov, D.~Solovyev, D.~Chubukov, S.~Chekhovskoi, and L.~Labzowsky, Phys. Rev. Res. \textbf{4}, L022052 (2022).
%
\bibitem{Sal58} E.~E.~Salpeter, Phys. Rev. \textbf{112}, 1642 (1958).
%
\bibitem{San65} P.~G.~H.~Sandars, Phys. Lett. \textbf{14}, 194 (1965).
%
\bibitem{Flam76} V.~V.~Flambaum, Yad. Fiz. \textbf{24}, 383 (1976)
[Sov. J. Nucl. Phys. \textbf{24}, 199 (1976)].
%
\bibitem{San67}
P.~G.~H.~Sandars, Phys. Rev. Lett. \textbf{19}, 1396 (1967).
%
\bibitem{Lab77}
L.~N.~Labzowsky, Zh. Eksp. Teor. Fiz. \textbf{73}, 1623 (1977) [Sov. Phys. JETP \textbf{46}, 853 (1977)].
%
\bibitem{Lab78}
L.~N.~Labzowsky, Zh. Eksp. Teor. Fiz. \textbf{75}, 856 (1978)
[Sov. Phys. JETP \textbf{48}, 434 (1978)]. 
%
\bibitem{San75}
P.~G.~H.~Sandars, At. Phys. \textbf{4}, 71 (1975).
%
\bibitem{Sush78}
O.~P.~Sushkov and V.~V.~Flambaum, Zh. Eksp. Teor. Fiz. \textbf{75}, 1208 (1978)
[Sov. Phys. JETP \textbf{48}, 608 (1978)].
%
\bibitem{Gor79}
V.~G.~Gorshkov, L.~N.~Labzowsky, and A.~N.~Moskalev, Zh. Eksp. Teor. Fiz. \textbf{76}, 414 (1979)
[Sov. Phys. JETP \textbf{49}, 209 (1979)].
%
\bibitem{Bon15}
A.~A.~Bondarevskaya, D.~V.~Chubukov, O.~Yu.~Andreev, E.~A.~Mistonova, L.~N.~Labzowsky, G.~Plunien, D.~Liesen, and F.~Bosch, J. Phys. B \textbf{48}, 144007 (2015).
%
\bibitem{Dmit87}
Yu.~Yu.~Dmitriev, M.~G.~Kozlov, L.~N.~Labzowsky, A.~V.~Titov, and V.~I.~Fomichev, J. Phys. B \textbf{20}, 4939 (1987).
%
\bibitem{Dmit92}
Yu.~Yu.~Dmitriev, Yu.~G.~Khait, M.~G.~Kozlov, L.~N.~Labzowsky, A.~O.~Mitrushenkov, A.~V.~Shtoff, and A.~V.~Titov, Phys. Lett. A \textbf{167}, 280 (1992).
%
\bibitem{Mos98}
N.~S.~Mosyagin, M.~G.~Kozlov, and A.~V.~Titov, J. Phys. B \textbf{31}, L763 (1998).
%
\bibitem{Is05}
T.~A.~Isaev, A.~N.~Petrov, N.~S.~Mosyagin, and A.~V.~Titov, Phys. Rev. Lett. \textbf{95}, 163004 (2005).
%
\bibitem{Tit06}
A.~V.~Titov, N.~S.~Mosyagin, A.~N.~Petrov, T.~A.~Isaev, and D.~P.~DeMille, Progr. Theor. Chem. Phys. \textbf{15}, 253 (2006).
%
\bibitem{Skrip09}
L.~V.~Skripnikov, A.~N.~Petrov, A.~V.~Titov, and N.~S.~Mosyagin, Phys. Rev. A \textbf{80}, 060501 (R) (2009).
%
\bibitem{Skrip16}
L.~V.~Skripnikov, J. Chem. Phys. \textbf{145}, 214301 (2016).
%
\bibitem{Koz95}
M.~G.~Kozlov and L.~N.~Labzowsky, J. Phys. B \textbf{28}, 1933 (1995).
%
\bibitem{Koz97}
M.~G.~Kozlov, J. Phys. B \textbf{30}, L607 (1997).
%
\bibitem{Quin98}
H.~M.~Quiney, H.~Skaane, and I.~P.~Grant, J. Phys. B \textbf{31}, L85 (1998).
%
\bibitem{Par98}
F.~A.~Parpia, J. Phys. B \textbf{31}, 1409 (1998).
%
\bibitem{Reg02}
B.~C.~Regan, E.~D.~Commins, C.~J.~Schmidt, and D.~DeMille, Phys. Rev. Lett. \textbf{88}, 071805 (2002).
%
\bibitem{Hud11}
J.~J.~Hudson, D.~M.~Kara, I.~J.~Smallman, B.~E.~Sauer, M.~R.~Tarbutt, and E.~A.~Hinds, Nature \textbf{473}, 493 (2011).
%
\bibitem{ACME13}
J.~Baron, et al. (ACME collaboration), Science \textbf{343}, 269 (2014).
%
\bibitem{ACME18}
V.~Andreev, et al. (ACME collaboration), Nature \textbf{562}, 355 (2018).
%
\bibitem{Cair17}
W.~B.~Cairncross, D.~N.~Gresh, M.~Grau, K.~C.~Cossel, T.~S.~Roussy, Y.~Ni, Y.~Zhou, J.~Ye, and E.~A.~Cornell, Phys. Rev. Lett. \textbf{119}, 153001 (2017).
%
\bibitem{JILA23}
 T.~S.~Roussy, et al. (JILA Collaboration), Science \textbf{381}, 46 (2023).
%
\bibitem{Liu92}
Z.~W.~Liu and H.~P.~Kelly, Phys. Rev. A \textbf{45}, R4210(R) (1992).
%
\bibitem{Dzuba09}
V.~A.~Dzuba and V.~V.~Flambaum, Phys.Rev. A \textbf{80}, 062509 (2009).
%
\bibitem{Por12}
S~.G.~Porsev, M.~S.~Safronova, and M.~G.~Kozlov, Phys. Rev. Lett. \textbf{108}, 173001 (2012).
%
\bibitem{Chub18}
D.~V.~Chubukov, L.~V.~Skripnikov, and L.~N.~Labzowsky, Phys. Rev. A \textbf{97}, 062512 (2018).
%
\bibitem{Quiney:98}
H.~M.~Quiney, H.~Skaane, and I.~P.~Grant, J. Phys. B: At. Mol. Opt. Phys.  \textbf{31}, 85 (1998).
%
\bibitem{Parpia:98}
F.~Parpia, J.Phys. B: At. Mol. Opt. Phys. \textbf{31}, 1409 (1998).
%
\bibitem{Skripnikov:13c}
L.~V.~Skripnikov, A.~N.~Petrov, and A.~V.~Titov, J. Chem. Phys. \textbf{139}, 221103 (2013).
%
\bibitem{Skripnikov:15a}
L.~V.~Skripnikov and A.~V.~Titov, J. Chem. Phys. \textbf{142}, 024301 (2015).
%
\bibitem{Skripnikov:16b}
L.~V.~Skripnikov, J. Chem. Phys. \textbf{145}, 214301 (2016).
%
\bibitem{Fleig:16}
M.~Denis and T.~Fleig, J. Chem. Phys. \textbf{145}, 214307 (2016).
%
\bibitem{Petrov:07a}
A.~N.~Petrov, N.~S.~Mosyagin, T.~A.~Isaev, and A.~V.~Titov, Phys. Rev. A \textbf{76}, 030501(R) (2007).
%
\bibitem{Skripnikov:17c}
L.~V.~Skripnikov, J. Chem. Phys. \textbf{147}, 021101 (2017).
%
\bibitem{Fleig:17}
T.~Fleig, Phys. Rev. A \textbf{96}, 040502 (2017).
%
\bibitem{Petrov:18b}
A.~N.~Petrov, L.~V.~Skripnikov, A.~V.~Titov, and V.~V.~Flambaum, Phys. Rev. A \textbf{98}, 042502 (2018).
%
\bibitem{Chub14}
D.~V.~Chubukov and L.~N.~Labzowsky, Phys. Lett. A \textbf{378}, 2857 (2014).
%
\bibitem{Chub17} D.~V~Chubukov and L.~N.~Labzowsky, Phys. Rev. A \textbf{96}, 052105 (2017).
%
\bibitem{Chub23} D.~V.~Chubukov, I.~A.~Aleksandrov, L.~V.~Skripnikov, and A.~N.~Petrov,
Phys. Rev. A \textbf{108}, 053103 (2023).
%
\bibitem{Chub19-1} D.~V.~Chubukov, L.~V.~Skripnikov, V.~N.~Kutuzov, S.~D.~Chekhovskoi, and L.~N.~Labzowsky, Atoms \textbf{7}, 56 (2019).
%
\bibitem{Chub19-2}
D.~V.~Chubukov, L.~V.~Skripnikov, and L.~N.~Labzowsky, JETP Letters {\bf 110}, 382 (2019).
%
\bibitem{Chub19-3} D.~V.~Chubukov, L.~V.~Skripnikov, L.~N.~Labzowsky, V.~N.~Kutuzov, and S.~D.~Chekhovskoi,
Phys. Rev. A {\bf 99}, 052515 (2019).
%
\bibitem{Chub21} 
D.~V.~Chubukov, L.~V.~Skripnikov, A.~N.~Petrov, V.~N.~Kutuzov, and L.~N.~Labzowsky,
Phys. Rev. A {\bf 103}, 042802 (2021).
%
\bibitem{Chekh} S.~D.~Chekhovskoi, D.~V.~Chubukov, L.~V.~Skripnikov, A.~N.~Petrov, and L.~N.~Labzowsky, Phys. Rev. A {\bf 108}, 052819 (2023).
%
\bibitem{Hoog90}
F.~Hoogeveen, Nucl. Phys. B \textbf{341}, 322 (1990).
%
\bibitem{Pos91}
M.~Pospelov and I.~Khriplovich, Yad. Fiz. \textbf{53}, 1030 (1991)
[Sov. J. Nucl. Phys. \textbf{53}, 638 (1991)].
%
\bibitem{Pos14}
M.~Pospelov and A.~Ritz, Phys. Rev. D \textbf{89}, 056006 (2014).
%
\bibitem{Yam21}
Y.~Yamaguchi and N.~Yamanaka, Phys. Rev. D \textbf{103}, 013001 (2021).
%
\bibitem{Ema22}
Y.~Ema, T.~Gao, and M.~Pospelov, Phys. Rev. Lett. \textbf{129}, 231801 (2022).
%
\bibitem{Chub16}
D.~V.~Chubukov and L.~N.~Labzowsky, Phys. Rev. A \textbf{93}, 062503 (2016).
%
\bibitem{LL4}
V.~B.~Berestetskii, E.~M.~Lifshitz, and L.~P.~Pitaevskii, 
{\it Quantum Electrodynamics} (Pergamon, Oxford, 1982).
%
\bibitem{Akhiez}
A.~I.~Akhiezer and V.~B.~Berestetskii, {\it Quantum Electrodynamics} (Wiley Interscience, New York, 1965). 
%
\bibitem{Cheng84}
T.-P.~Cheng and L.-F.~Li, {\it Gauge Theory of Elementary Particle Physics} (Clarendon Press, Oxford, 1984).
%
\bibitem{Shab78}
E.~P.~Shabalin, Yad. Fiz. \textbf{28}, 151 (1978)
[Sov. J. Nucl. Phys. \textbf{28}, 75 (1978)].
%
\bibitem{Khrip91}
I.~B.~Khriplovich, {\it Parity Nonconservation in Atomic Phenomena} (Gordon and Breach, London, 1991).
%
\bibitem{Jar85}
C.~Jarlskog, Phys. Rev. Lett. \textbf{55}, 1039 (1985).
%
\bibitem{Tan18}
M.~Tanabashi et al. (Particle Data Group), Phys. Rev. D \textbf{98}, 030001 (2018).
%
\bibitem{Ok82}
L.~B. Okun', {\it Leptons and Quarks} (North Holland, Amsterdam, 1982).
%
\end{thebibliography}
\end{document}